\documentclass{ptapap}
\usepackage{amsmath, amssymb}
\usepackage{bbold}

\author{Gerardo Urrutia}[CFT]
\author{Agnieszka Janiuk}[CFT]
\affil[CFT]{Center for Theoretical Physics, Polish Academy of Sciences, Al. Lotnikow 32/46, 02-668 Warsaw, Poland}

\title{The propagation of long GRB jets through and beyond its progenitor star}

\begin{document}

\maketitle

\begin{abstract}

Long gamma-ray bursts (lGRB) are produced by relativistic jets arising from the collapse of massive stars. Such progenitor environments present complex physical conditions that are challenging to model by numerical simulations. The difficulty increases when solving the accretion process and propagation of the outflows, as it requires covering distances from the black hole horizon to beyond the progenitor star. General Relativistic Magnetohydrodynamic (GRMHD) simulations provide a convenient framework to study high-luminosity jets, where magnetic flux plays an important role in the process of jet launching from the central engine. To follow the propagation of the jet through and beyond its progenitor environment, we use multi-scale simulations (i.e., AMR-based). In this work, we report results of 2.5-dimensional GRMHD simulations of a lGRB progenitor. We present highly magnetized, weakly magnetized, and non-magnetized pre-collapse stars, and discuss the observational implications for lGRB jets.

\end{abstract}

\section{Introduction}

The relativistic jets launched from collapsing massive stars produce long gamma-ray bursts (lGRBs) lasting $t \gtrsim 2$~s \citep[e.g.,][]{kumar15}. After the burst, a long-duration afterglow radiation is emitted from X-rays to radio frequencies \citep{Frailetal1997,Ramirez-Ruiz-Mcfadyen_2010,perley14}. The isotropic energy of a lGRB jet covers a range of $E_{\rm iso}\sim 10^{51}-10^{53}$~erg, in some cases even more energetic $E_{\rm iso}\sim 10^{54}$~erg \citep{Atteia2017,Angulo2024}. 

A black hole surrounded by an accretion disc (the central engine) launches the powerful jet \citep{paczynski91}. A fast-rotating massive stars ($M_\star \sim 25 M_\odot - 30 M_\odot$) must have magnetic fields $B_0\sim 10^{10}-10^{12}$~G \citep[e.g.,][]{MacFadyenWoosley1999,Burrows_2007}. Its collapse results in a high-spinning black hole $a\sim 0.9$. Such conditions are able to power the jet through \citet{bz77} mechanism and provide luminosity $L_{\rm BZ} \approx 10^{51}\left( M_{\rm BH}/5M_\odot \right)^2 \left( B_0 / 10^{15}{\rm G} \right)^2 \left( a/0.8 \right)^2\, {\rm erg}\,{\rm s}^{-1}$   \citep{MacFadyenWoosley1999,WoosleyBloom2006,KomissarovBarkov2009,BrombergTchekhovskoy2016,bugli2021,gottlieb2022a,Burrows2023,Vartanyan2025,Burrows2025,Morales-Rivera2025}.

In addition, high magnetic field strengths $B_0\sim 10^{14}-10^{16}$ can be sustained by the presence of a proto-neutron star (PNS). Then the magnetic flux is amplified, producing powerful magnetically driven jets \citep{Obergaulinger2006axisimetricMagneto,Obergaulinger2006A-2,Burrows_2007,Moesta2014,magnetorotational1,Shibata2024spinEv,FujibayashiShibata2025,Shibata2025arXiv,Urrutia2025Collapsar}.

The combined effects of accretion processes and the interaction of outflows with the progenitor environment alter the key jet properties. The jet can be additionally surrounded by a cocoon which modifies the picture of jet interaction \citep[e.g.,][]{decolle18b,Izzo2019Nature,Hamidani2023}. These effects have often been studied in isolation, either through accretion simulations or by imposing jet injection conditions by hand. Jet variability during the accretion process has been linked to characteristic times of Magneto Rotational Instabilities (MRI)  \citep[e.g.,][]{janiuk_variability2021}, and its potential repercussions on large scales by \citet{lopezcamara2016,harrison18,Becerra2025}, which aligns in the context of observed variability. During jet expansion, the conversion of the initial energy (whether dominated by magnetic or thermal components) into kinetic energy significantly impacts the jet and cocoon dynamics \citep{Tchekhovskoy2009,Matsumoto2019}. Combinations of these effects and their interactions with progenitor can be imposed by hand to extract the resulting jet properties, for example, how and where energy is dissipated, the breakout time, the final jet structure, and the cocoon's morphology \citep[e.g.,][]{urrutia22_3D,urrutia2025,Urrutia2025Collapsar}. A key open question is whether the results obtained from jets imposed at large scales can be consistently recovered and validated when the jet is launched self-consistently from the central engine \citep[e.g.,][]{gottlieb2022b,Gottlieb2023lowspining,GottliebSheGots2024}. 

In this study, we follow the propagation of a relativistic jet and the wind-driven outflows from the BH horizon and beyond the collapsing massive star. We perform 2.5-dimensional GRMHD simulations of three models that span different magnetization regimes.

\section{Methods}

We utilized \emph{BHAC} code \citep{Porth2016,Olivares2019a} to perform 2.5-dimensional GRMHD simulations. We use a logarithmically spaced grid in modified Kerr-Schild coordinates. The radial domain covers from the BH horizon up to $1\times 10^5$~$r_g$, while the azimuthal direction covers $\theta \in [0,\pi]~$rad. The Adaptive Mesh Refinement (AMR) has $3$ levels of refinement, employing $288$ cells in the $r$-direction, which increases to $1152$ at the maximum refinement level. In the $\theta$-direction, we use $64$ cells, increasing to $248$ at the highest refinement level. The maximum integration time $t_f$ employed in our simulations is $t_f \geq 2 \times 10^5$~$t_g$.

We obtained our lGRB progenitor by evolving a massive star with the \emph{MESA} code \citep{Paxton_2013,Paxton_2015,Paxton_2018,Paxton_2019}, starting from the main sequence up to the carbon burning stage. More details about this progenitor are provided in \citet{Urrutia2025Collapsar}. To follow the collapse of the progenitor, we remapped the density and pressure profiles from \textit{MESA} into our computational domain. The stellar mass is $M_\star = 25,M_\odot$ the radius $R_\star = 3.3 \times 10^{10}$~cm. We set a black hole mass of $M_{\mathrm{BH}} = 5,M_\odot$ at the centre. Considering the gravitational radius $r_g = GM_{\rm BH}/c^2$ and the gravitational time $t_g = r_g/c$, our computational domain covers distances of $\sim 10^{11}$~cm and the evolution time of $\sim 5$~s.

We assume a $\phi$-component of the velocity $u^\phi \sim C \sin^2\theta$, based on previous results from progenitor models \citep[e.g.,][]{WoosleyHeger_2006}, and we set $C = 2$ to guarantee the formation of a mini-disk in the evolution phase \citep{Murguiaberthier2020}. For the magnetic field geometry, we implemented dipole-like magnetic fields. We distinguish the models reported here as follows:
\begin{itemize}
    \item \textbf{Model 1:} It does not have a magnetic field.
    \item \textbf{Model 2:} We adopt the hybrid magnetic field utilized \citet{gottlieb2022a}, where inside the core of the star $\vec{B}=B_0~\hat{z}$, while outside is described by the potential $A_\phi=B_0\,\frac{r_c^3\sin\theta}{r} {\rm max}\left(\frac{r^2}{r^3+r_c^3}-\frac{R_\star^2}{R_\star^3+r_c^3},0 \right)\,$.
    \item \textbf{Model 3:} We implemented the dipole thought the $\phi$-component of the magnetic potential $A_\phi = B_0\, \frac{r_c^3}{r^3 + r_c^3}\, r \sin \theta\,$, where the stellar core is $r_c=10^8$~cm in both Model 2 and Model 3. \citep[See, ][]{Moesta2014,magnetorotational1}.
\end{itemize}
The magnetization $\sigma = B^2 / 4\pi \rho c^2$, depends on the density distribution of the star and the magnetic field geometry. We set the maximum value to $\sigma_{\rm max} \lesssim 0.2$. The magnetic field strengths considered in this work are below $B_0 \lesssim 10^{13}$~G. Such values are able to produce magnetic fluxes at the BH horizon sufficient to generate jets via the BZ mechanism suggested by \citet{Burrows_2007,KomissarovBarkov2009,gottlieb2022a}:
\begin{equation}
\Phi_{\rm BH,min} = 4 \pi r_h^2 |B_h| \approx 7 \times 10^{27}
\sqrt{\frac{\rho_{\rm max}^*}{10^7~{\rm g}~{\rm cm}^{-3}}}
,{\rm G},{\rm cm}^{-2}\,.
\label{eqn:mag_condition}
\end{equation}
In the next Section, we present the magnetic flux, jet luminosity, and breakout times produced by the three models during jet propagation.

\section{Results}

\begin{figure}
    \centering
    \includegraphics[width=0.49\textwidth]{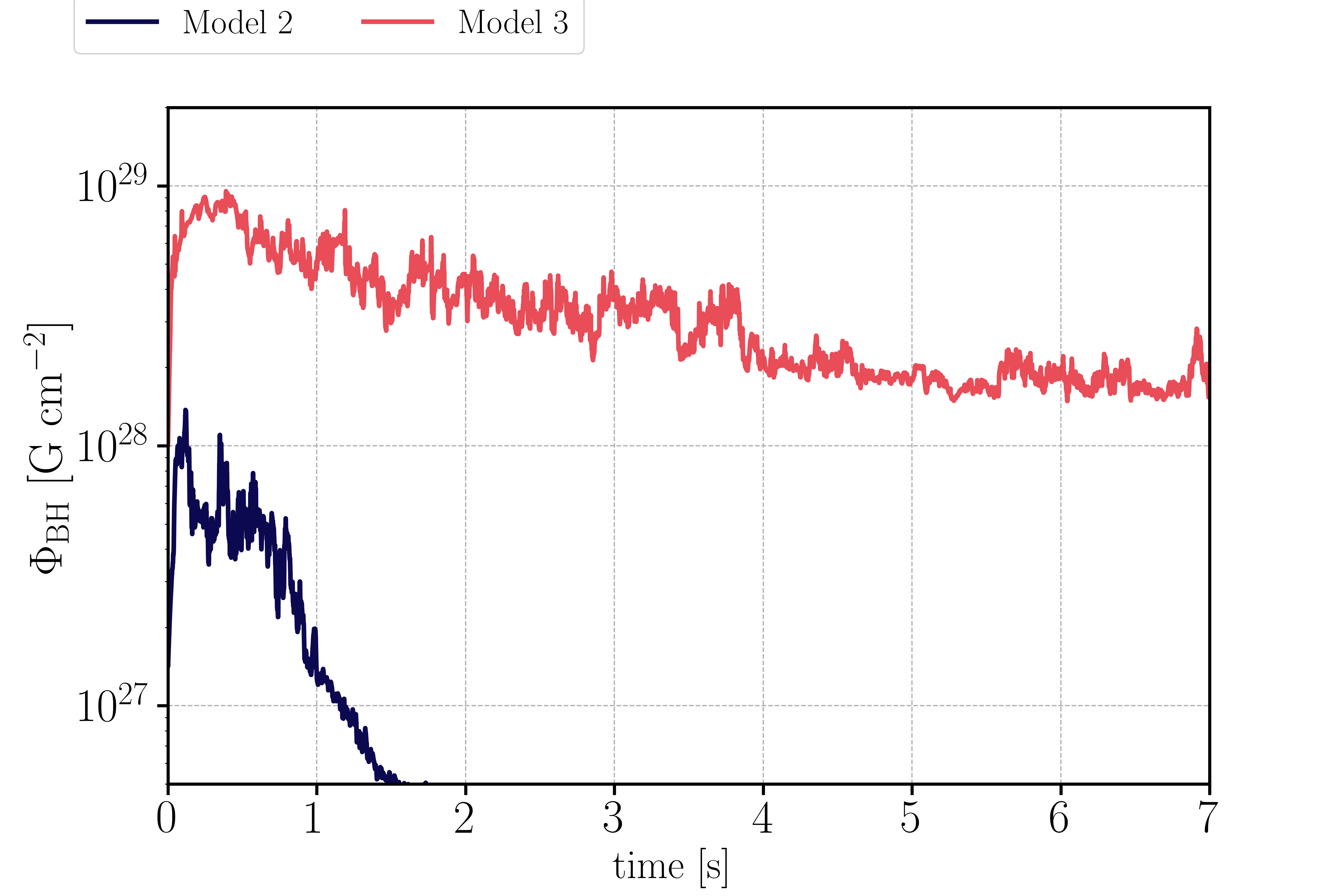}
    \includegraphics[width=0.49\textwidth]{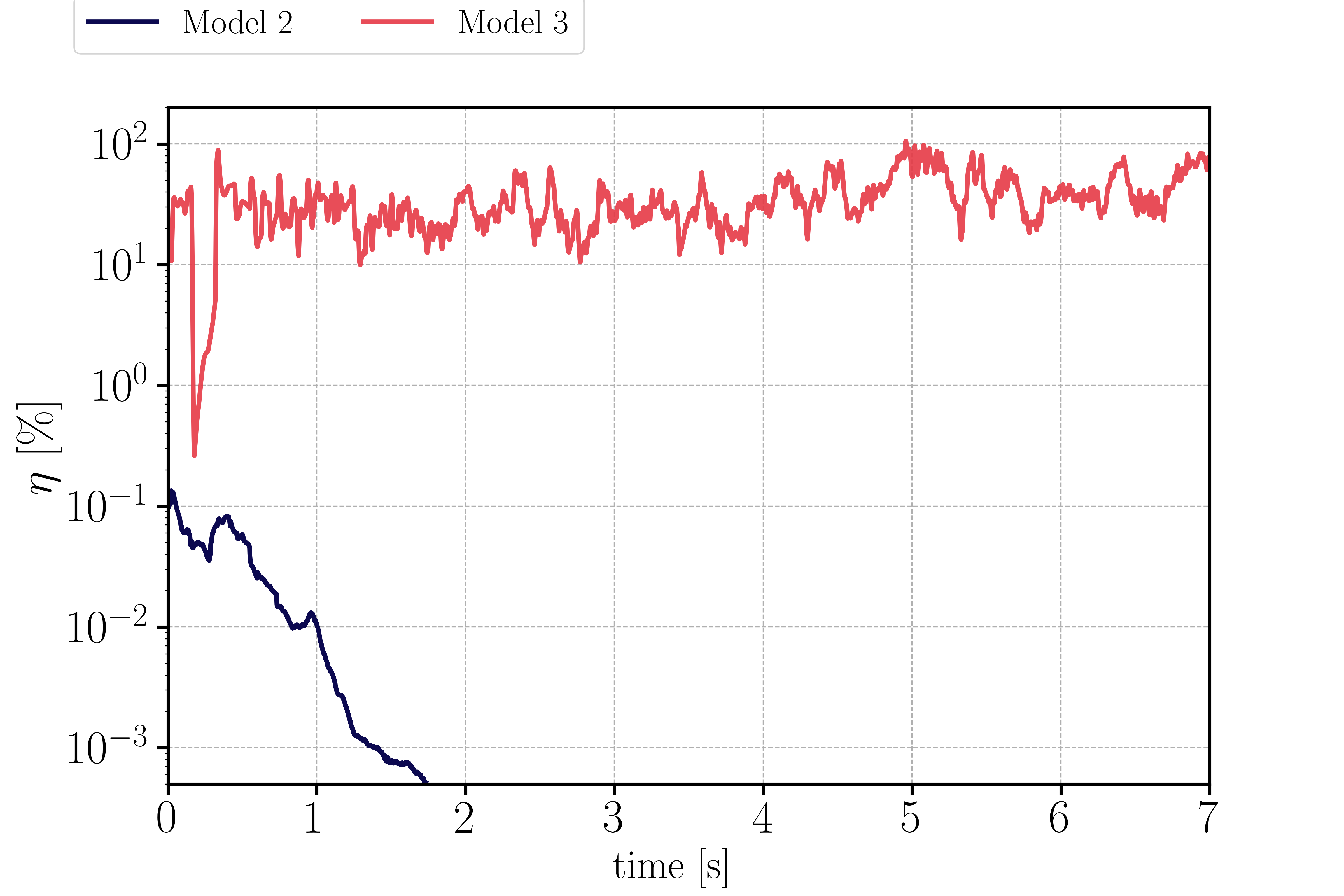}
    \caption{Left panel: the evolution of the magnetic flux. Right panel: the jet efficiency. Model 1 is not reported in this figure because it is not magnetized.}
    \label{fig:flux}
\end{figure}

The magnetic flux through a sphere of radius $r_{\rm sphere}=50r_g$, outside of BH horizon, is estimated as
%
    $\Phi_{\rm BH}=\frac{1}{2} \iint |B^{r}| \sqrt{-g} d\theta d\phi \; ,$
%
where $B^r$ is the radial component of the magnetic field and $g$ the determinant of the metric tensor $g_{\mu \nu}$. In the left panel of Figure~\ref{fig:flux}, for Model 2, the magnetic flux starts with values $\Phi_{\rm BH}\sim 10^{28}$~G~cm$^{-2}$ enough to drill the star satisfying the criterium given by eq.~\eqref{eqn:mag_condition}. However, it decreases fast during the first 1~s of evolution, meanwhile the model 3 present  a slow decay from $\Phi_{\rm BH}\sim 10^{29}$~G~cm$^{-2}$ to $\Phi_{\rm BH}\sim 10^{28}$~G~cm$^{-2}$ during $t\sim 7$~s. The electromagnetic component of the stress energy tensor is extracted as
%
    $L_{\rm EM} = -\frac{1}{4\pi} \int_{0}^{\pi/2} \int_{0}^{2\pi} \left(b^2 u^r u_t - b^r b_t \right) \sqrt{-g} \, d\phi \, d\theta \;,$
%
where $b^2=b^\mu b_\mu$, with $b^\mu$ is the co-moving four-vector of the contravariant magnetic field in the fluid frame, and $u^{\mu}$ the four-vector of the velocity. The accretion rate is estimated by
%
    $\dot{M}=-\iint \rho c^2 u^r \sqrt{-g} d\theta d\phi \;$,
%
where both $L_{\rm EM}$ and $\dot{M}$ are measured outside of the BH horizon, at $r_{\rm sphere}=50r_g$. Then, the jet efficiency is
%
    $\eta = L_{\rm EM}/\dot{M}c^2 \;.$
%
This is shown in the right panel of Figure~\ref{fig:flux}. The Model 3, which is magnetically saturated, reaches efficiencies of $\eta > 50\%$. This magnetic energy is extracted via the Blandford–Znajek process, i.e., the jet luminosity exceeds the energy rate of the infalling material \citep{McKinney2005,Salafia2021AGiacomazzo2021}. On the other hand, the weak jet remains in the regime $\eta < 0.01\%$.

\begin{figure}[!h]
    \centering
    \includegraphics[width=\textwidth]{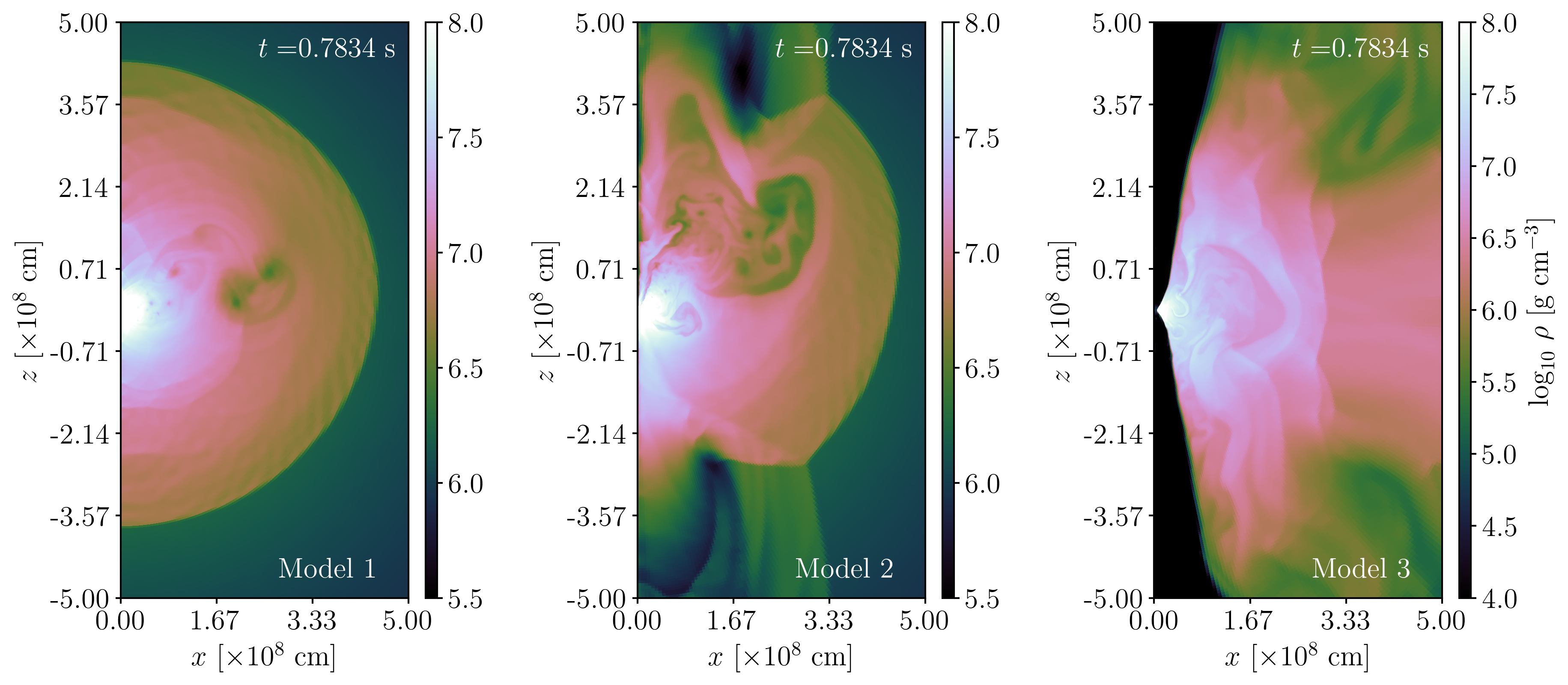}
    \includegraphics[width=\textwidth]{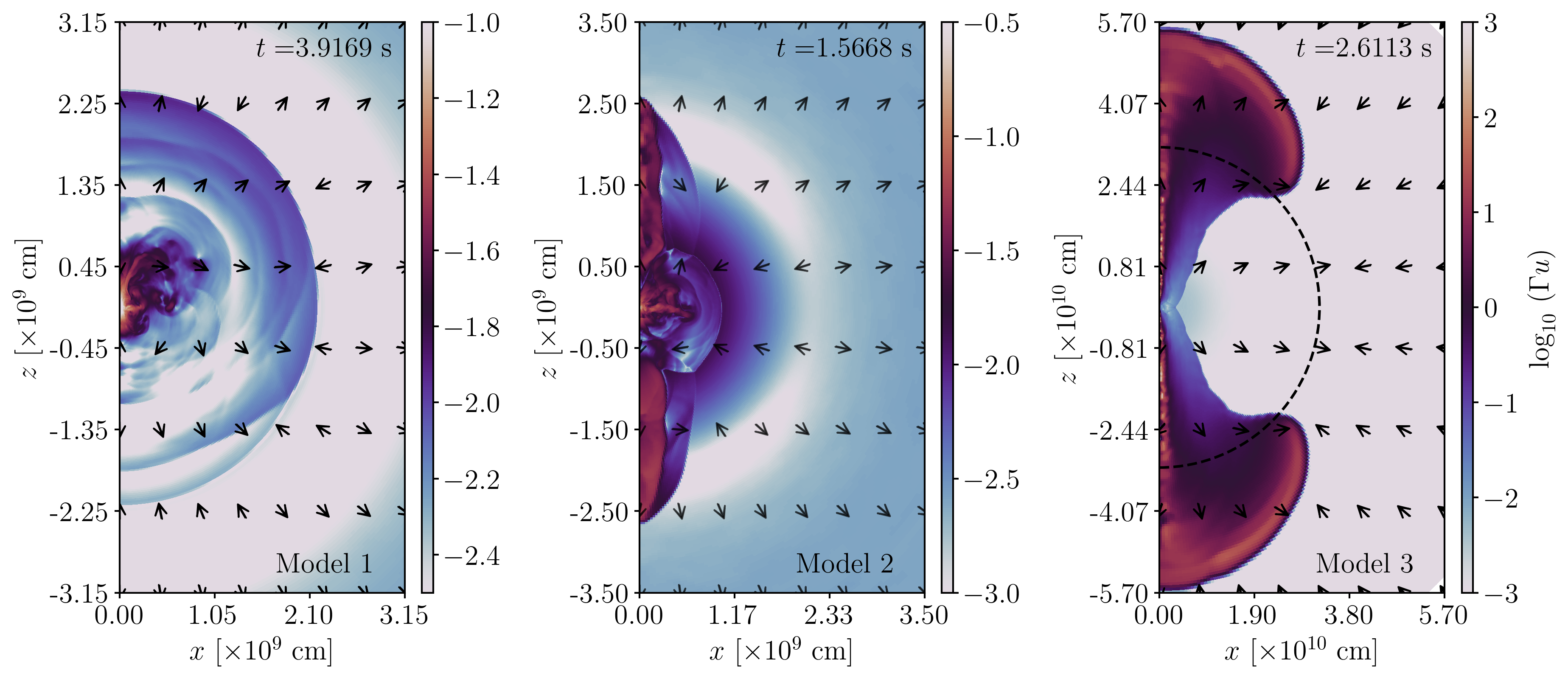}
    \caption{Upper row: the density maps zoomed into the region of the stellar core $r \sim 10^8$ cm. We show the same snapshot at $t \sim 0.78$~s to compare the ejected outflows by the accretion process. Lower row: maps of the velocity magnitude $\Gamma u$. We show at different times and different zoom levels. The dashed circle in Model 3 represents the stellar boundary.}
    \label{fig:maps}
\end{figure}

In Figure~\ref{fig:maps}, we present snapshots of the performed models at different times. The upper panel shows the density $\rho$ distribution at core scales $r\sim 10^8$~cm, which is the densest region of the star. In the lower panel, we present the magnitude of the four velocity $\Gamma u$. The accretion disk encloses the highest-density material near the centre. The magnetic flux has a significant influence on the dynamics of each model, particularly in terms of jet launching, which in turn alters the density distribution around the central engine. 

The Model 1 in the first column of Figure~\ref{fig:maps} shows the accretion process in the absence of magnetic fields. The density distribution shows the expected formation of an accretion disk, consistent with initial angular momentum parameter $C=2$ \citep[See,][]{Murguiaberthier2020}. However, a low-density funnel along the $z$-axis is not formed. The jet will not be powered due to the absence of any flux at the centre (magnetic or neutrino).  

The Model 2 in the middle column of Figure~\ref{fig:maps} shows the propagation of a weak jet. Jet formation occurs more slowly, which breaks the dense bubble of accreting material surrounding the black hole. While some high-density material remains concentrated in the centre, the magnetic flux diminishes rapidly. As a result, the jet lacks sufficient flux to accelerate inside the surrounding ram pressure and is likely to fail, remaining confined within the dense core material. The velocity map presents a maximum $\Gamma u \sim 2$, which in turn could break the star in several seconds if the magnetic flux remains close to $\Phi_{\rm BH}\sim 10^{27}$~G~cm$^{-2}$.

The Model 3 exhibits a successfully launched jet. A well-defined jet region forms at the centre, surrounded by dense material. As the jet propagates, the shock interacts with the stellar core, redistributing the dense core material and pushing it laterally. The Poynting flux continues to drive the shock front to large distances, extending into the less dense outer layers of the star and beyond. The velocity map presents values of $\Gamma u \sim 10^3$ in the jet region together with variability. The cocoon material also accelerates once it propagates outside of the progenitor.

\section{Discussion and Conclusions}


In this work, we present the evolution of 2.5-dimensional GRMHD simulations of collapsars, with azimuthal velocity structure $u^\phi \sim C \sin^2\theta$ and different magnetization. The progenitor star has a mass of $M_\star = 25 M_\odot$, at its centre we impose a black hole with spin $a = 0.9$ and mass $M_{\rm BH} = 5 M_\odot$. We first explore the evolution of Model 1, which is not magnetized. Although the accretion disk region is visible and the funnel appears two orders of magnitude less dense than the disk, this contrast is insufficient to classify it as a low-density funnel. This result confirms that an additional energy flux is required to produce such a structure, as the magnetized models successfully open a jet funnel through the action of magnetic flux.

Model 3 exhibits a persistent magnetic flux able to drive a successful jet launched via the Blandford–Znajek (BZ) mechanism. It also shows a high jet efficiency, consistent with a magnetically arrested disk (MAD) state. In contrast, Model 2 presents a high magnetic flux at early times and decays abruptly after two seconds of evolution. In addition, its jet efficiency remains very low, then it does not reach the MAD state, limiting the extraction of rotational energy from the BH.

Model 1 produces a strong accretion-disk wind. However, after $t = 2$ s, the shock front stops its expansion (see, for example, Figure~8 in~\citet{urrutia2025}). As expected, sustained wind expansion requires a continuous thermal energy source \citep[e.g.,][]{Crosato2024,Crosato2025}. Assuming a constant luminosity and a total injected energy of $E \sim 10^{52}$ erg for a 12TH progenitor star (different progenitor star than adopted in this proceeding) with radius $R_\star \approx 9 \times 10^{10}$ cm, \citet{Urrutia2023Gws} find a breakout time of $t_{\rm bo} = 62$ s, while a failed jet yields $t_{\rm bo} = 45$ s.  In our Model 2, the failed jet also stops its expansion due to the absence of an additional energy flux. In contrast, Model 3 exhibits a much shorter breakout time of $t_{\rm bo} = 2.2$ s and a high jet luminosity of $L_j = 5 \times 10^{52}$ erg s$^{-1}$. Similar short breakout times have been previously reported for highly luminous jets \citep[e.g.,][]{Hamidani2017,Urrutia2023Gws}. Jets with luminosities $L_j \lesssim 10^{52}$ erg s$^{-1}$ typically show longer breakout times, in the range $t \sim 5$–$10$ s \citep[e.g.,][]{lopezcamara2009,lopezcamara2016,Hamidani2017,harrison18,Hamidani2021-expanding,Suzuki2022,urrutia22_3D,Pais2023collapsars}.


The structure of the progenitor star plays a crucial role in the formation and propagation of jets. In this work, we consider only a single progenitor model. However, in \citet{Urrutia2025Collapsar} we report jet formation and propagation for alternative progenitors, namely: the 12TH and 16TI performed by \citet{WoosleyHeger_2006}. We find that the conversion of jet energy into kinetic energy depends sensitively on the progenitor’s structure. In some cases, this leads to the production of flares, driven by rapid energy conversion when the density drops abruptly beyond the stellar core. A set of progenitor models has also been explored using semi-analytical approaches, demonstrating how progenitor structure affects jet luminosities and breakout times \citep{Morales-Rivera2025}.


To connect our results with observations, we associate the origin of variability in the early prompt emission with the magnetorotational instabilities (MRI), which induce fluctuations in the magnetic flux and the Lorentz factor. It modifies the photospheric emission \citep[e.g.,][]{lopezcamara2016}. In addition, the progenitor star influences the final jet structure $E(\theta)$, leading to observable modifications in the afterglow emission. This effect has been estimated based on collapsar simulations of jets propagating from the BH horizon to large scales \citep[e.g.,][]{Urrutia2025Collapsar}.

This study is limited by its spatial dimension. In some cases, it has been shown that axisymmetric (2D) setups can successfully launch jets, whereas fully three-dimensional simulations do not \citep[e.g.,][]{Moesta2014}. On the other hand, both 2D and 3D studies have revealed qualitatively similar behavior, with the main differences arising from the level of environmental perturbations \citep[e.g.,][]{magnetorotational1}. In our case, the results are expected to be comparable, since the magnetic flux in our models is of the same order as that reported in 3D studies by \citet{gottlieb2022a,gottlieb2022b}, suggesting that our jets would also form in three dimensions. However, our simulations do not capture inherently three-dimensional effects such as jet wobbling or disk tilting. In addition, the black hole mass is constant over time. A dedicated study exploring lower black hole spins \citep[e.g.,][]{Gottlieb2023lowspining}, while carefully accounting for spin equilibrium, is therefore required \citep[e.g.,][]{Issa2025b}. Nevertheless, the scenario presented here is consistent with rapidly rotating progenitors, which are plausible sources of high-luminosity lGRBs or super-luminous supernovae \citep[e.g.,][]{Aguilera-Dena2018ApJ}. The absence of neutrino transport in our models prevents us from fully reproducing jet breakout in the low-luminosity case and from accurately modeling the shock front driven by the strong wind of the non-magnetized accretion disk \citep[e.g.,][]{Moesta2014,magnetorotational1,Issa_2025,Janiuk2025NeutrinoLeakeage}. Finally, the evolution of the black hole spin is not included in this study (c.f., \citealt{Janiuk2023collapse}; Płonka \& Janiuk 2025 submitted).

\acknowledgements{This work was supported by the grant 2023/50/A/ST9/00527 from Polish National Science Center. We gratefully acknowledge Polish high-performance computing infrastructure PLGrid
(HPC Center: ACK Cyfronet AGH) for providing computer facilities and support within computational grant no. PLG/2025/018086.}

\bibliographystyle{ptapap}
\bibliography{urrutia}

\end{document}